# STATUS AND PLANS FOR A SRF ACCELERATOR TEST FACILITY AT FERMILAB*

J. Leibfritz[†], R. Andrews, K. Carlson, B. Chase, M. Church, E. Harms, A. Klebaner, M. Kucera, S. Lackey, A. Martinez, S. Nagaitsev, L. Nobrega, P. Piot, J. Reid, M. Wendt, S. Wesseln, FNAL, Batavia, IL 60510, USA

## Abstract

A superconducting RF accelerator test facility is being constructed at Fermilab. The existing New Muon Lab (NML) building is being converted for this facility. The accelerator will consist of an electron gun, injector, beam acceleration section consisting of 3 TTF-type or ILC-type cryomodules, multiple downstream beam lines for testing diagnostics and conducting various beam tests, and a high power beam dump. When completed, it is envisioned that this facility will initially be capable of generating an 810 MeV electron beam with ILC beam intensity. Expansion plans of the facility are underway that will provide the capability to upgrade the accelerator to a total beam energy of 1.5 GeV. In addition to testing accelerator components, this facility will be used to test RF power equipment, instrumentation, LLRF and controls systems for future SRF accelerators such as the ILC and Project-X. This paper describes the current status and overall plans for this facility.

## INTRODUCTION

Fermi National Accelerator Laboratory is constructing a superconducting radio frequency (SRF) accelerator test facility in the existing New Muon Lab (NML) building. The goal is to build a state-of-the-art complex for testing SRF cryomodules with beam for the development of next generation high intensity linear accelerators, such as the International Linear Collider (ILC) and Project-X.

Initially conceived as a test area for the ILC R&D program, the objective was to build a freestanding electron linac, capable of achieving the ILC goal of testing one RF Unit with ILC beam intensity [1]. An ILC RF Unit consists of three ILC-type cryomodules, each containing eight high gradient 9-cell 1.3 GHz niobium SRF cavities, powered by a single 10 MW pulsed RF system. The design beam parameters for this facility are shown in Table 1.

Table 1: Beam Parameters

| Bunch Charge | 3.2 nC |
| --- | --- |
| Pulse Length | 1 msec |
| Bunches Per Pulse | 3000 |
| Repetition Rate | 5 Hz |
| Bunch Length | < 300 μm RMS |
| Injector Beam Energy | 40 MeV |
| Final Beam Energy (1 RF Unit) | 810 MeV |

## FACILITY DESCRIPTION

The test facility consists of a shielding cave which houses the accelerator, as well as areas for the support equipment and systems required for operation (Fig. 1). The accelerator consists of three distinct sections: the Injector, SRF Accelerator, and the Test Beam Lines.

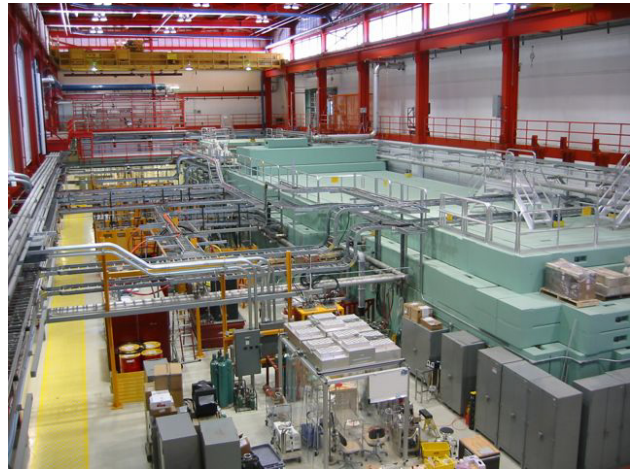

Figure 1: SRF Accelerator Test Facility at NML

### Injector

The injector is comprised of a 1.3 GHz photocathode electron RF gun, which generates the electron beam [2]. Following the gun are two cryomodule cryostats called Capture Cavities I & II (CC1 & CC2). Each of these contains a single high-gradient 1.3 GHz 9-cell superconducting RF cavity. These devices provide the initial acceleration of the electrons to approximately 40 MeV. The remainder of the injector is comprised of a bunch compressor and a series of magnets and beam diagnostics used to achieve the proper beam parameters before entering the string of SRF accelerating cryomodules.

### SRF Accelerator

The accelerator portion consists of a series of 12-meter long SRF cryomodules. The first cryomodule (CM1) is a TTF Type-III+ design, containing eight 1.3 GHz 9-cell superconducting RF cavities (Fig. 2). This style of cryomodule was developed by the TESLA collaboration

---

and was provided to Fermilab as a "Kit" from DESY. It is the first cryomodule of this type in the U.S. and was assembled by DESY, Fermilab, and INFN [3]. The full RF Unit test will include a series of three TTF-type or ILC-type cryomodules capable of generating a beam of approximately 810 MeV.

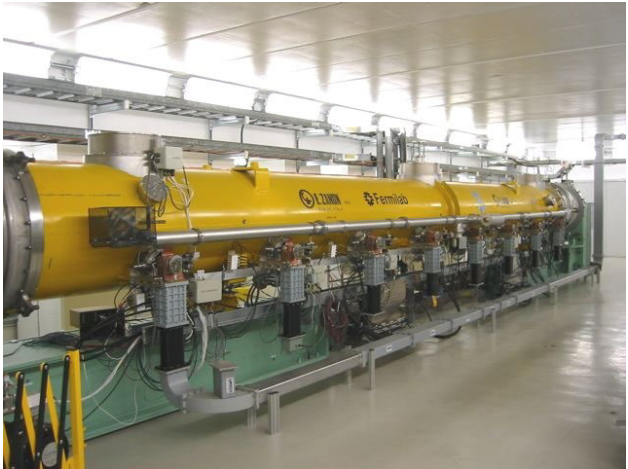

Figure 2: First installed Cryomodule (CM1)

### Test Beam Lines

Downstream of the accelerating cryomodules is the test beam line section. This area consists of an array of multiple high-energy beam lines that transport the electron beam from the accelerating cryomodules to one of three beam absorbers. Each absorber is capable of dissipating up to 75 kW of power and is surrounded by a large steel and concrete shielding dump. These beam lines are intended for various beam studies and R&D testing of future accelerator components and instrumentation.

### Support Systems

In addition to testing the accelerator components, the intent of this facility is to also test the support systems required for a future SRF linac. These systems include RF power, low-level RF, controls, instrumentation, low conductivity cooling water, and cryogenics. The facility has also been instrumented with two sophisticated vibration measurement systems that measure ground motion movements over time, as well as vibrations and accelerations of the SRF cavities in real-time [4-5].

## STATUS AND SCHEDULE

### Current Status

The construction of this facility is a large undertaking, which is being done in multiple phases. The first phase included the installation of the infrastructure necessary to operate the facility: cryogenics, water, power, RF, and controls, as well as building the test cave. Once the infrastructure was in place, CC2 (containing a single 9-cell cavity) and CM1 (containing eight 9-cell cavities) were installed, cooled down to 2 K (23.4 Torr), and RF power tested. This phase of the project has recently been completed with the cool down of CM1 in November 2010 [6] and the subsequent RF powering of the cavities beginning in December 2010.

### Next Phases

The next phase of the project involves preparing everything needed to generate the first beam in the accelerator. This includes: the installation and commissioning of the electron gun and associated laser system; completion of the injector portion of the accelerator; installation of the test beam lines; and fabrication and installation of the high energy beam absorbers. This phase is currently underway with a plan for first beam in 2012.

Following the generation of the first beam, the subsequent phases of the project will involve adding additional cryomodules to the accelerator section to complete the full RF Unit for testing.

## EXPANSION AND FUTURE PLANS

Expansion plans are underway that involve substantial civil construction. These plans focus on expanding the length of the accelerator tunnel, upgrading the cryogenic system, and providing an area to test various types of cryomodules. In addition to the SRF cryomodule tests, there are also future plans for testing low-energy 3.9 GHz crab-cavities; conducting advanced accelerator R&D of future accelerator components using the high-energy beam lines; and testing of 325 MHz and 650 MHz cryomodules.

### Tunnel Expansion

The first construction project is the addition of a 70 m-long tunnel to the existing NML facility. This expansion essentially doubles the length of the test accelerator from 75 m (the length of the existing NML building) to 140 m. This will provide enough space for up to six (12 m-long) cryomodules and increase the beam energy capability of the facility from 810 MeV to 1.5 GeV. In addition to providing additional length to the overall accelerator, the expansion also includes a large 15 m-wide area for the high-energy test beam lines at the downstream end of the accelerator; as well as space to install a 10 m-diameter storage ring to conduct future accelerator R&D programs and experiments. An enclosure to house the high-energy beam absorbers and dump is situated at the end of the test beam lines. Construction began in March 2010 and is expected to be complete in April 2011.

### New Cryogenic Plant

In the early stages of operation of the test facility, the cryogenics for the SRF cavities are provided by a pair of Tevatron-style satellite refrigerators in combination with a warm vacuum pumping system. This system is capable of providing approximately 120 W of super fluid helium, which is not sufficient to support the full operation of the facility [6-7]. Therefore, a new cryogenics plant is being added to the facility to supplement the existing system's capacity. The new cryogenics system will be capable of supporting simultaneous operation of a linac with up to six SRF cryomodules, as well as various modes of operation of future cryomodule test stands.

*Cryomodule Test Facility*

A pair of adjoining buildings called the Cryomodule Test Facility (CMTF) is being constructed adjacent to the existing NML facility. CMTF will house the new cryogenics plant as well as multiple stand-alone SRF cryomodule test stands. One of the buildings will house the noisy vibrating equipment (compressors, pumps, etc.) needed to operate the cryoplant. The other building will contain: the cryogenic cold boxes, cryomodule test stands, RF systems, a vacuum clean room, and an office area. The current plan is for this facility to house two test caves that are capable of testing various styles of cryomodules at 325 MHz, 650 MHz and 1.3 GHz, in pulsed and continuous wave (CW) modes of operation. The construction of CMTF began in October 2010 and is expected to be complete by the end of 2011. A layout of the entire SRF Accelerator Test Facility complex, including these expansion projects, is shown in Figure 3.

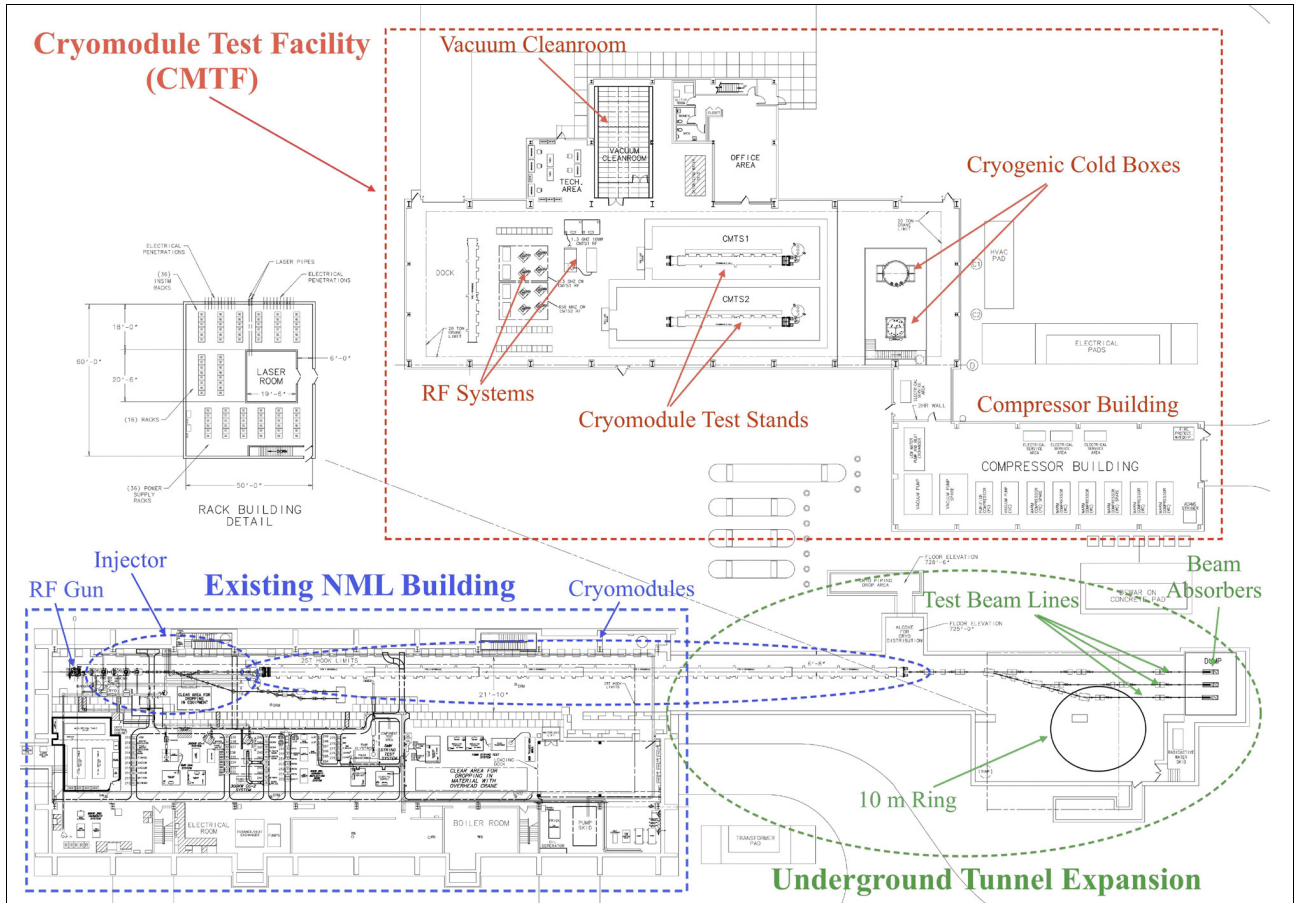

Figure 3: Layout of SRF Accelerator Test Facility

## ACKNOWLEDGEMENTS

Authors would like to acknowledge technical specialists C. Exline, D. Franck, W. Johnson, R. Kellett and C. Rogers for all their efforts, as well as the entire NML Project Team.